\documentclass{conm-p-l}

\copyrightinfo{2008}{}

\usepackage{graphicx}
\usepackage{subfigure}

\usepackage{pstricks}
\usepackage{pst-plot}

\newtheorem{theorem}{Theorem}[section]

\theoremstyle{definition}

\theoremstyle{remark}
\newtheorem{remark}[theorem]{Remark}

\numberwithin{equation}{section}

\newcommand{\tu}{\tilde{u}}

\newcommand{\hk}{\hat{k}}

\newcommand{\e}{\epsilon}

\newcommand{\C}{{\mathcal C}}

\newcommand{\Ga}{\Gamma}

\newcommand{\dl}{\delta}
\newcommand{\Dl}{\Delta}
\renewcommand{\th}{\theta}
\newcommand{\ra}{\rightarrow}
\newcommand{\al}{\alpha}

\newcommand{\sg}{\sigma}
\newcommand{\Sg}{\Sigma}

\newcommand{\pa}{\partial}

\newcommand{\la}{\lambda}

\newcommand{\nid}{\noindent}

\newcommand{\om}{\omega}
\newcommand{\Om}{\Omega}
\newcommand{\na}{\nabla}

\newcommand{\lra}{\longrightarrow}

\renewcommand{\O}{{\mathcal O}}

\newcommand{\non}{\nonumber}

\newcommand{\Z}{\mathbb{Z}/\{0\}}
\newcommand{\ZZ}{\mathbb{Z}^2/\{0\}}

\newcommand{\hu}{\hat{u}}
\newcommand{\tom}{\tilde{\omega}}

\begin{document}

\title[Chaos Phenotypes]{Chaos Phenotypes in Fluids}

\author{Y. Charles Li}
\address{Department of Mathematics, University of Missouri, 
Columbia, MO 65211, USA}
\curraddr{}
\email{cli@math.missouri.edu}
\thanks{}


\subjclass{Primary 37, 76; Secondary 35, 34}
\date{}

\dedicatory{}


\begin{abstract}
I shall briefly survey the current status on more rigorous studies of chaos in 
fluids by focusing along the line of chaos phenotypes: sensitive dependence 
on initial data, and recurrence.
\end{abstract}

\maketitle

\tableofcontents










\section{Introduction}

Chaos has two distinctive phenotypes: sensitive dependence on initial data, 
and recurrence. Sensitive dependence on initial data means that no matter 
how small the initial condition changes, after sufficiently long time, the 
change will reach order one. This phenotype can also be observed in non-chaotic 
systems, for example, in an explosive system. Together with the second phenotype,
it can often identify chaos. Recurrence means that the orbit repeatedly re-visits 
the neighborhood of its initial point. 

Sensitive dependence on initial data can often be proved in a neighborhood of a homoclinic 
orbit or a heteroclinic cycle. The best technique for accomplishing this is the so-called 
shadowing lemma \cite{Li04}. On the other hand, recurrence can often be established from 
general measure-theoretic arguments. 

There are several good prototypes for understanding chaos in fluids: 2D Navier-Stokes equations
under periodic boundary conditions and periodic forcing in both space and time, the Couette flow,
the Poiseuille flow, and the plane Poiseuille flow, to name a few. When the so-called Reynolds number 
is large enough, all these flows develop chaos. When the Reynolds number just crosses its threshold,
the chaos is often transient (that is, it has a finite life time). With the increase of the 
Reynolds number, the life time of chaos increases too. Transient chaos is very poorly understood even 
for ordinary differential equations. The rigorously proved chaos is always eternal. Sometimes, one can 
observe intermittent chaos in fluids, that is, the chaos state and the regular state repeatedly 
exchange in time to eternity. 

In order to identify the chaos phenotypes in fluids, one can try to prove a recurrence theorem 
for Euler equations \cite{Li08a} \cite{Li08b}, and establish the existence of a heteroclinic cycle 
\cite{LL08}. As a chaos phenotype, the recurrence theorem proved in \cite{Li08a} is quite 
satisfactory, while the heteroclinic cycle search in \cite{LL08} is still at a conjecture stage.  
Another relevant topic to the dynamical system study of fluids is invariant manifold. It turns out 
that an unstable manifold can be proved for 2D Navier-Stokes equations, while invariant manifolds 
for Euler equations are a tough open problem \cite{Li05}. All these studies were conducted on the 
simplest prototype --- 2D Navier-Stokes equations under periodic boundary conditions. 
For other prototypes mentioned above, very little is understood about chaos. Their analytical studies 
were often focused upon linear instability. There is a famous Sommerfeld paradox saying that 
Couette flow is linearly stable for all Reynolds numbers as first calculated by Sommerfeld 
\cite{Som08}, but experiments show that Couette flow is unstable when the Reynolds number is 
large enough. This paradox lies at the heart of understanding turbulence inside the infinite 
dimensional phase space. An analytical study on the paradox is conducted in \cite{LL08a}.
Another study was given in \cite{KM08}. A complete resolution of the paradox will require a 
complete understanding of the dynamics at least in a certain region of the infinite 
dimensional phase space. On the other hand, investigating the paradox may lead the way and have 
significant impact in the studies of fluid dynamics in phase spaces. 
Recently, there has been a renaissance in numerical dynamical system studies on the paradox \cite{OP80} \cite{Nag90} \cite{KK01} \cite{Wal03} \cite{Ker05} \cite{Vis07} \cite{Eck08} \cite{VC08} \cite{GHC08}. New non-wandering solutions (fixed points, periodic orbits, and relative periodic orbits) are discovered. Better estimates on the critical Reynolds number and perturbation threshold are obtained. It seems that pipe Poiseuille flow, plane Couette flow, and plane Poiseuille flow share some universal coherent structure of ``streak, roll, and critical layer''. The coherent structure seems quite robust in the sense that they survive for increasing Reynolds number. There seems to be also good agreement with experiments \cite{Hof04}. In the infinite Reynolds number limit, the ``streak'' part of the fixed points tends to a limit shear (one of the infinitely many shear fixed points of 3D Euler equations); and the ``roll'' and ``critical layer'' disappear \cite{WGW07} \cite{Vis08}. The limit shear is characterized by a condition \cite{Li08}. 

The article is organized as follows. In section 2, we shall focus on recurrence. In sections 3-7,
we shall discuss the issues related to sensitive dependence on initial data, especially the 
heteroclinics conjecture, linear spectra and their zero-viscosity limits, and the Melnikov 
integral. In section 8, we shall briefly discuss the Sommerfeld paradox.
Section 9 is a discussion on future directions and open problems.

\section{Recurrence}

\begin{figure}[ht]
\includegraphics[width=4.0in,height=3.0in]{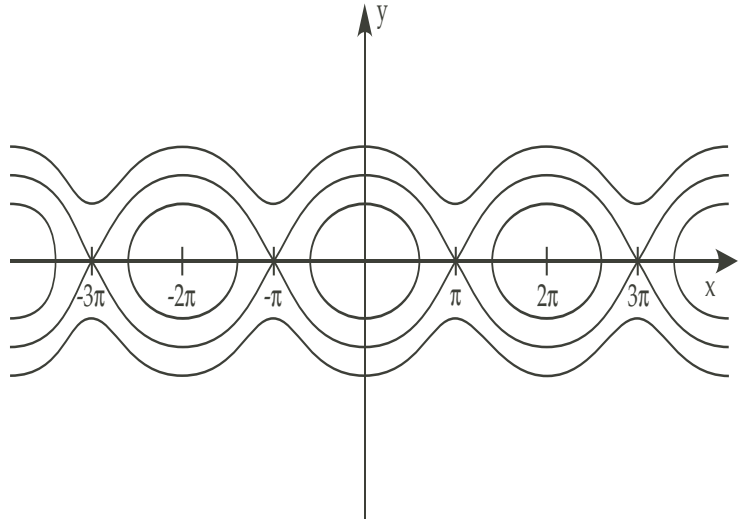}
\caption{The phase plane diagram of the pendulum equation.}
\label{ppd}
\end{figure}

The classical Poincar\'e recurrence theorem can be stated as follows:
\begin{theorem}[Poincar\'e Recurrence Theorem]
Let ($X, \Sg , \mu$) be a finite measure space and $f\ : \ X \mapsto X$ be a 
measure-preserving transformation. For any $E \in \Sg$ ($\sg$-algebra of 
subsets of $X$), the measure
\[
\mu (\{ x \in E \ | \ \exists N,\  f^n(x) \not\in E \ \forall n > N \} ) = 0 \ .
\]
That is, almost every point returns infinitely often.
\end{theorem}
The geometric intuition of the Poincar\'e 
recurrence theorem is that in a finite measure space (or invariant subset), the 
images of a positive measure set under a measure-preserving map will have no room left but 
intersect the original set repeatedly. The measure of the space $X$ being finite is crucial. 
For example, consider the two-dimensional Hamiltonian system of the pendulum
\begin{equation}
\dot{x} = y \ , \quad \dot{y} = - \sin x \ . 
\label{pex}
\end{equation}
Its phase plane diagram is shown in Figure \ref{ppd}. If the invariant region 
includes orbits outside the cat's eyes, then the measure of the region will not 
be finite, and the Poincar\'e recurrence theorem will not hold. One can see clearly 
that the orbits outside the cat's eyes will drift to infinity. 

The investigation of recurrence of fluid dynamics in an infinite dimensional phase space
encounters a serious problem that natural finite dimensional measures (e.g. Gibbs measure) 
do not have good counterparts in infinite dimensions. It is well-known that the 
kinetic energy and enstrophy are invariant under the 2D Euler flow. But it is difficult to 
use them to define finite measures in infinite dimensions. It seems possible to study 
the Poincar\'e recurrence problem directly from Banach norms rather than measures. Along 
this direction, the concept of measure-preserving transformation, which is so critical in finite 
dimensions, is not necessary anymore. 

Consider the 2D Euler equation
\[
\pa_t u+(u\cdot \na )u = -\na p, \quad \na \cdot u = 0 ;
\]
where $u=(u_1,u_2)$ is the velocity of the fluid, and $p$ is the pressure.
The boundary conditions are crucial in determining the structure of the dynamics in the 
phase space. The simplest boundary condition is the periodic boundary condition, that is, 
topologically the fluid motion is confined on the 2-torus $\mathbb{T}^2$. Under such a boundary 
condition, certain recurrence can be established as discussed below \cite{Li08a}. The 2D 
Euler equation is globally well-posed in $H^s(\mathbb{T}^2)$ ($s>2$). We also require that 
\begin{equation}
\int_{\mathbb{T}^2} u \ dx = 0.
\label{mzc}
\end{equation}
The theorem stated below says that any $H^s(\mathbb{T}^2)$ ($s>2$) solution to the 2D Euler 
equation returns repeatedly to an arbitrarily small $H^0(\mathbb{T}^2)$ neighborhood.
\begin{theorem} \cite{Li08a}
For any $\tu \in H^s(\mathbb{T}^2)$ ($s>2$), any $\dl >0$, and any $T>0$; there is a $u^* 
\in H^s(\mathbb{T}^2)$ such that
\[
F^{m_jT}(\tu ) \in B^0_\dl (u^*)=\{ \hu \in H^s(\mathbb{T}^2) \  | \  \| \hu -u^* \|_{H^0(\mathbb{T}^2)}
< \dl \}
\]
where $\{ m_j \}$ is an infinite sequence of positive integers, and $F^t$ is the evolution operator 
of the 2D Euler equation.
\label{TR}
\end{theorem}

Periodic boundary condition makes fluid dynamics a more natural dynamical system. The hope is that 
certain conclusions derived from the periodic boundary condition may be universal among all  
boundary conditions, or at least in the region away from boundaries. Of course, there are 
always special properties that are unique to the 
specific boundary conditions, for example, the example in \cite{Li08b} \cite{Nad91} of non-return 
near initial point is mainly due to the special slip boundary conditions on the boundaries of 
an annulus. 

\section{Heteroclinics Conjecture}

2D Euler equation is an infinite dimensional Hamiltonian system \cite{Arn66}. Under periodic 
boundary condition, its Hamiltonian structure can be written in terms of Fourier series 
\cite{Li00}. In a Hamiltonian system, heteroclinics often comes from the so-called separatrix. 
Such a separatrix represents the instability in the Hamiltonian system. The general scenario 
of chaos generation is that when the Hamiltonian system is under proper perturbations, chaos 
is created near the separatrix. Proper perturbations often represent a good balance between 
energy dissipation and energy input. For instances, going from 2D Euler equation to 2D 
Navier-Stokes equation, proper perturbations can be obtained by proper external forcings, 
e.g. spatially and temporally periodic forcing \cite{LL08}. 

For the heteroclinics search, it is more convenient to write the 2D Euler equation in the vorticity 
form,
\begin{equation}
\pa_t \Om + \{ \Psi, \Om \} = 0,
\label{2DEV}
\end{equation}
where $\Om$ is the vorticity which is a real scalar-valued function
of three variables $t$ and $x=(x_1, x_2)$, the bracket $\{\ ,\ \}$ 
is defined as
\[
\{ f, g\} = (\pa_{x_1} f) (\pa_{x_2}g) - (\pa_{x_2} f) (\pa_{x_1} g) \ ,
\]
where $\Psi$ is the stream function given by,
\[
u_1=- \pa_{x_2}\Psi \ ,\ \ \ u_2=\pa_{x_1} \Psi \ ,
\]
the relation between vorticity $\Om$ and stream 
function $\Psi$ is,
\[
\Om =\pa_{x_1} u_2 - \pa_{x_2} u_1 =\Dl \Psi \ ,
\]
We pose the periodic boundary condition
\[
\Om (t, x_1 +2\pi , x_2) = \Om (t, x_1 , x_2) = \Om (t, x_1, x_2 +2\pi /\al ),
\]
where $\al$ is a positive constant, i.e. the 2D Euler equation is defined on the 
2-torus $\mathbb{T}^2$. 

We propose the following conjecture \cite{LL08}.
\begin{itemize}
\item The Heteroclinics Conjecture: In the Sobolev space 
$H^\ell (\mathbb{T}^2)$ ($\ell \geq 3$), for any fixed point $\Om$ of the 2D Euler 
flow having an unstable eigenvalue, there is a pair of heteroclinic cycles asymptotic 
to the two fixed points 
$\Om$ and $-\Om$.
\end{itemize}

The nature of ``a pair of heteroclinic cycles'' is motivated from the following symmetries:
\begin{enumerate}
\item $\Om (t, x_1, x_2) \lra \Om (t, -x_1, -x_2)$,
\item $\Om (t, x_1, x_2) \lra -\Om (-t, x_1, x_2)$,
\item $\Om (t, x_1, x_2) \lra -\Om (t, -x_1, x_2)$, or $\Om (t, x_1, x_2) \lra -\Om (t, x_1, -x_2)$,
\item $\Om (t, x_1, x_2) \lra \Om (t, x_1+\th_1, x_2+\th_2)$, $\quad \forall \th_1, \th_2$.
\end{enumerate}
The first symmetry allows us to work in an invariant subspace in which all the $\om_k$'s are 
real-valued. This corresponds to the cosine transform in (\ref{FS}). Take $\Om = \Ga \cos x_1$ 
as an example, the second symmetry maps the unstable manifold (assuming its existence) of the 
fixed point $\Ga \cos x_1$ into the stable manifold (assuming its existence) of $-\Ga \cos x_1$. 
The third symmetry maps the unstable manifold of $\Ga \cos x_1$ into the unstable manifold of 
$-\Ga \cos x_1$. By choosing $\th_1 =\pi$, the fourth symmetry maps the unstable manifold of 
$\Ga \cos x_1$ into the unstable manifold of $-\Ga \cos x_1$. To maintain the cosine transform, 
the $\th_1$ and $\th_2$ in the fourth symmetry can only be $\pi$ and $\pi /\al$. 
If there is a heteroclinic orbit asymptotic to $\Ga \cos x_1$ 
and $- \Ga \cos x_1$ as $t \ra -\infty$ and $+\infty$, then there may be two corresponding to the 
unstable eigenvector and its negative. In fact, both may lie on certain sphere in the phase 
space due to the constraint by the invariants. Then the third symmetry generates 
another pair of heteroclinic orbit asymptotic to $-\Ga \cos x_1$ and $\Ga \cos x_1$ as 
$t \ra -\infty$ and $+\infty$. Together they form a pair of heteroclinic cycles.

Using the Fourier series
\begin{equation}
\Om = \sum_{k \in \ZZ} \om_k e^{i(k_1 x_1 + \al k_2 x_2)}\ , 
\label{FS}
\end{equation}
where $\om_{-k} = \overline{\om_k}$, one gets the kinetic form of the 2D Euler equation
\[
\dot{\om}_k = \sum_{k=m+n} A(m,n) \ \om_m \om_n \ ,
\]
where 
\[
A(m,n) = \frac{\al}{2}\left [ \frac{1}{n_1^2+(\al n_2)^2} - 
\frac{1}{m_1^2+(\al m_2)^2}\right ]\left | \begin{array}{lr} 
m_1 & n_1 \\ m_2 & n_2 \\ \end{array} \right | \ .
\]

Denote by $\Sg$ the hyperplane
\[
\Sg = \left \{ \om \ | \ \om_k = 0 \ , \quad \forall \text{ even } k_2 \right \} \ .
\]
We have the following theorem.
\begin{theorem}\cite{LL08}
Assume that the fixed point $\Om = \Ga \cos x_1$ has a 1-dimensional local unstable manifold $W^u$, 
and $W^u \cap \Sg \neq \emptyset$; then the heteroclinics conjecture is true, i.e.  
there is a pair heteroclinic cycles to the 2D Euler equation that connects 
$\Om = \Ga \cos x_1$ and $-\Om$.
\label{hcthm}
\end{theorem}
Notice that the existence of invariant manifolds around the fixed point $\Om =\Ga \cos x_1$ 
is an open problem.

\section{Numerical Verification of the Heteroclinics Conjecture}

Besides the symmetries mentioned in last section, we will also make use of the 
conserved quantities: kinetic energy $E=\sum |k|^{-2} \om_k^2$ 
(where $|k|^2=k_1^2+\al^2 k_2^2$) and enstrophy $S=\sum \om_k^2$, which 
will survive as conserved quantities for any symmetric Galerkin truncation, to help us to 
track the heteroclinic orbit. We will only consider the case that all the $\om_k$'s are 
real-valued (i.e. $\cos$-transform). 

\begin{figure}
\includegraphics[width=4.0in,height=3.0in]{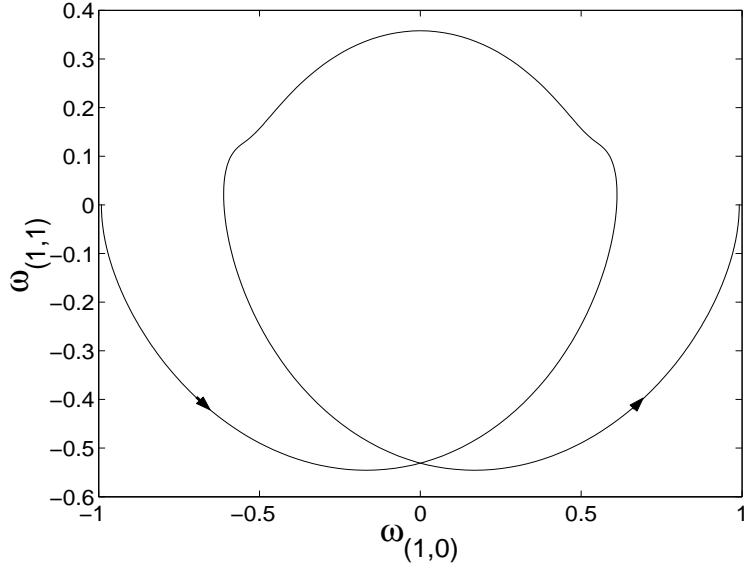}
\caption{The approximate heteroclinic orbit projected onto the ($\om_{(1,0)},\om_{(1,1)}$)-plane
in the case of the $\{ |k_1| \leq 2, |k_2| \leq 2 \}$ Galerkin truncation of the 2D Euler 
equation.}
\label{fi1}
\end{figure}

We make a Galerkin truncation by keeping modes: $\{ |k_1| \leq 2, |k_2| \leq 2 \}$, which results
in a $12$ dimensional system. We choose $\al =0.7$. After careful consideration of the above 
mentioned symmetries and conserved quantities ($E=S=1$), we discover the following initial 
condition that best tracks the heteroclinic orbit:
\begin{eqnarray}
& & \om_{(j,0)}=\om_{(j,2)} = 0 \ , \quad \forall j\ , \non \\
& & \om_{(0,1)}=0.603624\ , \quad \om_{(1,1)}=- \om_{(-1,1)}=0.357832\ , \label{IC} \\
& & \om_{(2,1)}=\om_{(-2,1)}=0.435632\ . \non 
\end{eqnarray}
We used fourth-order Runge-Kutta scheme. We also tested even higher-order Runge-Kutta schemes 
which do not improve the accuracy too much. 
Starting from this initial condition, we calculate the solution in both forward and backward time
for the same duration of $T =11.8$, and we discover the approximate heteroclinic orbit 
asymptotic to $2\cos x_1$ and $-2\cos x_1$ as $t \ra -\infty$ and $+\infty$, as shown in Figure 
\ref{fi1}. Then the third symmetry generates another heteroclinic orbit asymptotic to 
$-2\cos x_1$ and $2\cos x_1$ as $t \ra -\infty$ and $+\infty$. Together they form a heteroclinic 
cycle. Finally the second symmetry generates another heteroclinic cycle. That is, we have 
a pair of heteroclinic cycles. Notice also that the approximate heteroclinic orbit in 
Figure \ref{fi1} has an extra loop before landing near $-2\cos x_1$. This is due to the 
$k_2=2$ modes in the Galerkin truncation. For smaller Galerkin truncations, the heteroclinic orbits 
can be calculated exactly by hand and have no such extra loop \cite{Li03e} \cite{Li06e}, 
and existence of chaos generated by the heteroclinic orbit can be rigorously proved in some 
case \cite{Li06e}.
\begin{remark}
We have also conducted numerical experiments on Galerkin truncations by keeping more modes: 
$\{ |k_1| \leq 4, |k_2| \leq 4 \}$ and $\{ |k_1| \leq 8, |k_2| \leq 8 \}$. We found orbits 
that have similar behavior as the approximate heteroclinic orbit in Figure \ref{fi1}, but 
their approximations to heteroclinics are not as good as the one in Figure \ref{fi1}. 
We also tested the heteroclinics conjecture for models of 2D Navier-Stokes equation \cite{LL08}.
\end{remark}

\section{Linear Instability and Invariant Manifold}

The 2D Navier-Stokes equation in the vorticity form can be written as 
\begin{equation}
\pa_t \Om + \{ \Psi, \Om \} = \e [\Dl \Om + f(t,x)] ,
\label{2DNS}
\end{equation}
where $\e = 1/R$ is the inverse of the Reynolds number, we will consider the same boundary 
condition as for (\ref{2DEV}), $f(t,x)$ is the forcing,
\[
\int_{\mathbb{T}^2} f dx = 0 . 
\]

For the external force $f= \Ga \cos x_1$, $\Om = \Ga \cos x_1$ is a shear fixed 
point, where $\Ga$ is an arbitrary real nonzero constant. Choose $\al \in (0.5, 0.84)$. There is a 
$\e_* > 0$ such that when $\e > \e_*$, the fixed point has no eigenvalue 
with positive real part, and when $\e \in [0, \e_*)$, the fixed point has 
a unique positive eigenvalue \cite{Li05}. 
Notice that this unique eigenvalue persists even for linear Euler 
($\e =0$). In fact, for linear Euler ($\e =0$), there is a pair of 
eigenvalues, and the other one is the negative of the above eigenvalue. 
Precise statements on such results are given in the theorem below. Using the Fourier series 
(\ref{FS}) where we work in the 
subspace where all the $\om_k$'s are real-valued, we get
the spectral equation of the linearized 2D Navier-Stokes operator at the 
fixed point $\Om = 2 \cos x_1$, 
\begin{equation}
A_{n-1} \om_{n-1} -\e |\hk +np|^2 \om_n - A_{n+1} \om_{n+1} = \la \om_n \ ,
\label{le}
\end{equation}
where $\hk \in \ZZ$, $p=(1,0)$, $\om_n = \om_{\hk +np}$, $A_n = A(p, \hk +np)$, and
\[
A(q,r) = \frac{\al}{2}\left [ \frac{1}{r_1^2+(\al r_2)^2} - 
\frac{1}{q_1^2+(\al q_2)^2}\right ]\left | \begin{array}{lr} 
q_1 & r_1 \\ q_2 & r_2 \\ \end{array} \right | \ .
\]
(In fact, the $A_n$'s should be counted twice due to switching $q$ and $r$, but the 
difference is only a simple scaling of $\e$ and $\la$.)
Thus the 2D linear NS decouples according to lines labeled by $\hk$. The following 
detailed theorem on the spectrum of the 2D linear NS at the fixed point 
$\Om = 2 \cos x_1$ was proved in \cite{Li05}.
\begin{theorem}[The Spectral Theorem \cite{Li05}]
The spectra of the 2D linear NS operator (\ref{le}) have the following 
properties.
\begin{enumerate}
\item $(\al \hk_2)^2+(\hk_1+n)^2 > 1$, $\forall n \in \Z$. 
When $\e > 0$, there is no eigenvalue of non-negative real part. 
When $\e = 0$, the entire spectrum is the continuous spectrum
\[
\left [ -i\al |\hk_2|, \ i\al |\hk_2| \right ]\ .
\]
\item $\hk_2 = 0$, $\hk_1 = 1$. The spectrum consists of the eigenvalues 
\[
\la = - \e n^2 \ , \quad n \in \Z \ .
\]
The eigenfunctions are the Fourier modes
\[
\tom_{np} e^{inx_1} + \ \mbox{c.c.}\ \ , \quad \forall \tom_{np} \in 
\C\ , \quad n \in \Z \ .
\]
As $\e \ra 0^+$, the eigenvalues are dense on the negative half of the real 
axis $(-\infty, 0]$. Setting $\e =0$, the only eigenvalue is $\la = 0$ of 
infinite multiplicity with the same eigenfunctions as above.
\item $\hk_2 = -1$, $\hk_1 = 0$. (a). $\e >0$. For any $\al \in (0.5, 0.95)$,
there is a unique $\e_*(\al)$,
\begin{equation}
\frac{\sqrt{32-3\al^6-17\al^4-16\al^2}}{2(\al^2+1)(\al^2+4)} < 
\e_*(\al) < \frac{1}{(\al^2+1)} \sqrt{\frac{1-\al^2}{2}}\ ,
\label{nuda}
\end{equation}
where the term under the square root on the left is positive for 
$\al \in (0.5, 0.95)$, and the left term is always less than the right term.
When $\e > \e_*(\al)$, there is no eigenvalue of non-negative real part. 
When $\e = \e_*(\al)$, $\la =0$ is an eigenvalue, and all the rest 
eigenvalues have negative real parts. When $\e < \e_*(\al)$, there is 
a unique positive eigenvalue $\la (\e )>0$, and all the rest 
eigenvalues have negative real parts. $\e^{-1} \la (\e )$ is a strictly 
monotonically decreasing function of $\e$. When $\al \in (0.5, 0.8469)$,
we have the estimate
\begin{eqnarray*}
& & \sqrt{\frac{\al^2(1-\al^2)}{2(\al^2+1)}-\frac{\al^4 (\al^2+3)}{4
(\al^2+1)(\al^2+4)}} - \e (\al^2+1) < \la (\e ) \\
& & < \sqrt{\frac{\al^2(1-\al^2)}{2(\al^2+1)}}- \e \al^2 \ ,
\end{eqnarray*}
where the term under the square root on the left is positive for 
$\al \in (0.5, 0.8469)$.
\[
\sqrt{\frac{\al^2(1-\al^2)}{2(\al^2+1)}-\frac{\al^4 (\al^2+3)}{4
(\al^2+1)(\al^2+4)}} \leq \lim_{\e \ra 0^+} \la (\e )  \leq 
\sqrt{\frac{\al^2(1-\al^2)}{2(\al^2+1)}} \ .
\]
In particular, as $\e \ra 0^+$, $\la (\e ) =\O (1)$.

(b). $\e =0$. When $\al \in (0.5, 0.8469)$, we have only two eigenvalues
$\la_0$ and $-\la_0$, where $\la_0$ is positive,
\[
\sqrt{\frac{\al^2(1-\al^2)}{2(\al^2+1)}-\frac{\al^4 (\al^2+3)}{4
(\al^2+1)(\al^2+4)}} < \la_0 <
\sqrt{\frac{\al^2(1-\al^2)}{2(\al^2+1)}} \ .
\]
The rest of the spectrum is a continuous spectrum $[-i\al , \ i\al ]$.

(c). For any fixed $\al \in (0.5, 0.8469)$,
\begin{equation}
\lim_{\e \ra 0^+} \la (\e ) = \la_0 \ .
\label{pet1}
\end{equation}
\item Finally, when $\e = 0$, the union of all the above pieces of 
continuous spectra is the imaginary axis $i\mathbb{R}$.
\end{enumerate}
\label{PET}
\end{theorem}
Based upon the above spectral theorem, the following invariant manifold theorem 
can be proved.
\begin{theorem}[Invariant Manifold Theorem \cite{Li05}]
For any $\al \in (0.5, 0.95)$, and $\e \in (0, \e_*(\al ))$ where 
$\e_*(\al ) > 0$ satisfies (\ref{nuda}), in a neighborhood of $\Om = 2 \cos x_1$ 
in the Sobolev space $H^\ell (\mathbb{T}^2)$ ($\ell \geq 3$),
there are an $1$-dimensional $C^\infty$ unstable manifold and an 
$1$-codimensional $C^\infty$ stable manifold.
\end{theorem}

\section{Zero-Viscosity Limit of the Linear Spectra}

One of the goals of the work \cite{Li05} is to study the zero viscosity limit of 
the invariant manifolds of the 2D NS. For this study, it is crucial to understand 
the deformation of the linear spectra as $\e \ra 0^+$. Of course, studying this 
limit is of great interest in its own right. It is an interesting but difficult 
analysis problem too. We have conducted some numerical studies. We truncate 
(\ref{le}) at different sizes and compute the eigenvalues of the resulting matrices. 
We increase the truncation size until we see reliability of the result. We also tested 
the continued fraction approach \cite{Li00} for computing eigenvalues, the result is 
much worse. So we dropped the continued fraction approach. 

We find that 
the spectra of the linear Navier-Stokes and Euler operators can be classified into four 
categories in the zero viscosity limit \cite{LL08}: 
\begin{enumerate}
\item {\em Persistence:} These are the eigenvalues that persist and approach 
to the eigenvalues of the corresponding linear Euler operator when the viscosity 
approaches zero. (e.g. at 2D and 3D shears, and cat's eye.)
\item {\em Condensation:} These are the eigenvalues that approach and form 
a continuous spectrum for the corresponding linear Euler operator when the viscosity 
approaches zero. (e.g. at 2D and 3D shears, cat's eye, and ABC flow.)
\item {\em Singularity:} These are the eigenvalues that approach to a 
set that is not in the spectrum of the corresponding linear Euler operator when 
the viscosity approaches zero. (e.g. at 2D and 3D shears.)
\item {\em Addition:} This is a subset of the spectrum of the linear Euler operator,
which has no overlap with the zero viscosity limit set of the spectrum of the linear 
NS operator. (e.g. cat's eye.)
\end{enumerate} 

When $\hk_1=0$ and $\hk_2=1$, $\al =0.7$, the unique $\e_*$ in (\ref{nuda}) 
belongs to the interval $0.332 < \e_* < 0.339$,
such that when $\e < \e_*$, a positive eigenvalue appears. We test this criterion 
numerically and find that it is very sharp even when the truncation of 
the linear system (\ref{le}) is as low as $|n| \leq 100$. 
As $\e \ra 0^+$, we tested the truncation of 
the linear system (\ref{le}) up to $|n| \leq 1024$ for $\al =0.7$, the patterns are 
all the same. Below we present the case $|n| \leq 200$ for which the pattern is more clear
\cite{LL08}. 

\begin{figure}[ht] 
\centering
\subfigure[$\e =0.14$]{\includegraphics[width=2.3in,height=2.3in]{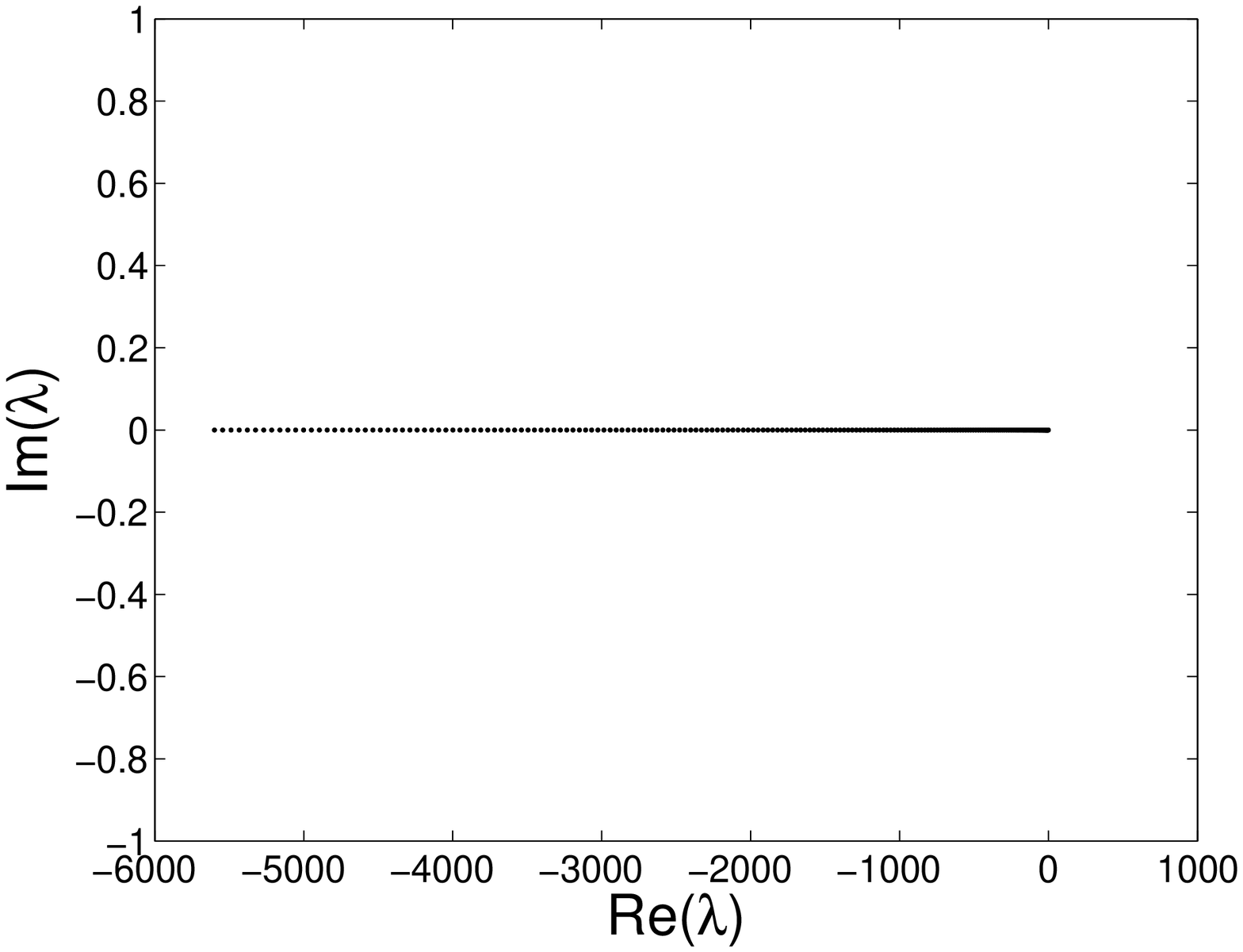}}
\subfigure[$\e =0.13$]{\includegraphics[width=2.3in,height=2.3in]{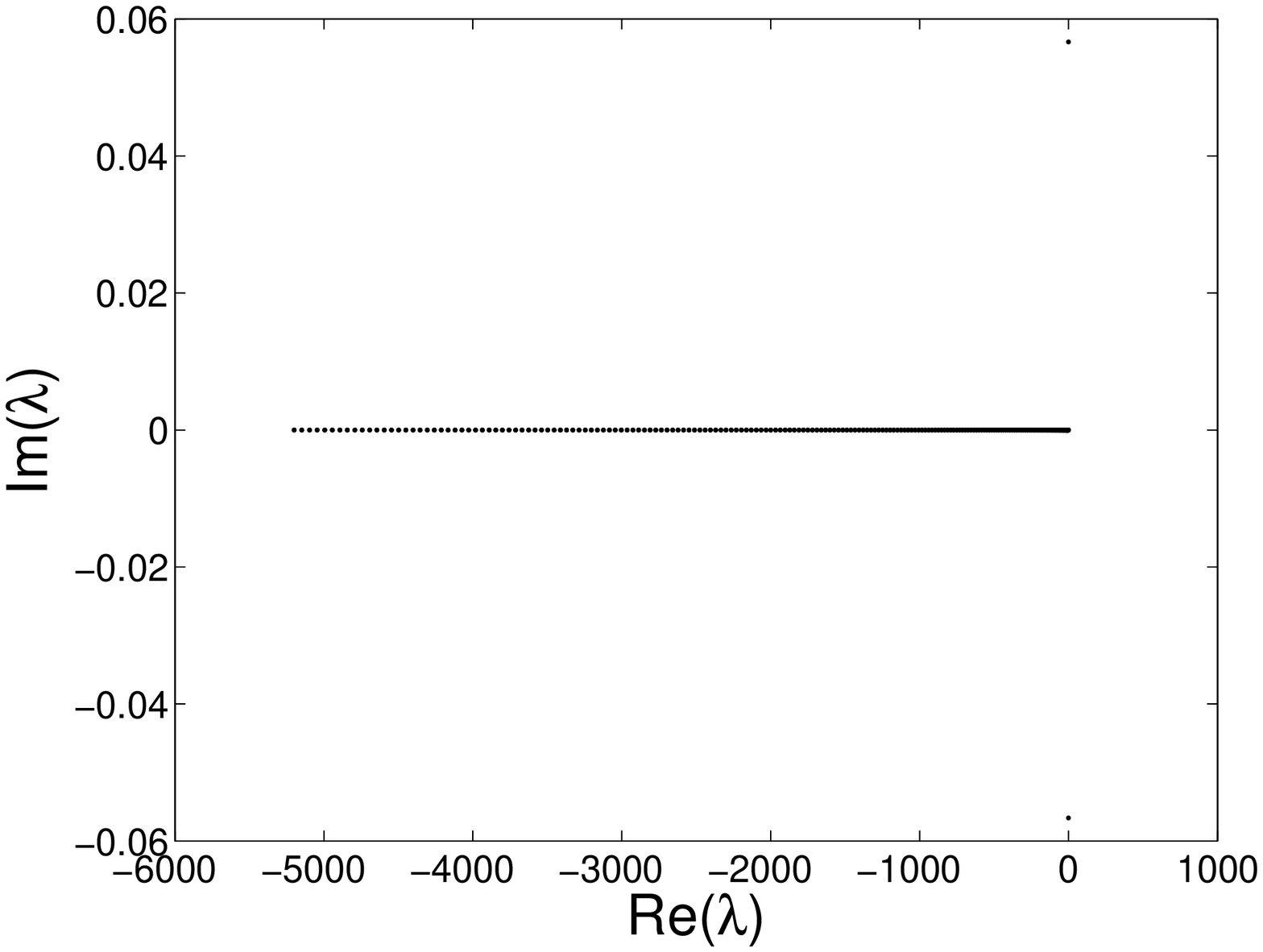}}
\subfigure[$\e =0.07$]{\includegraphics[width=2.3in,height=2.3in]{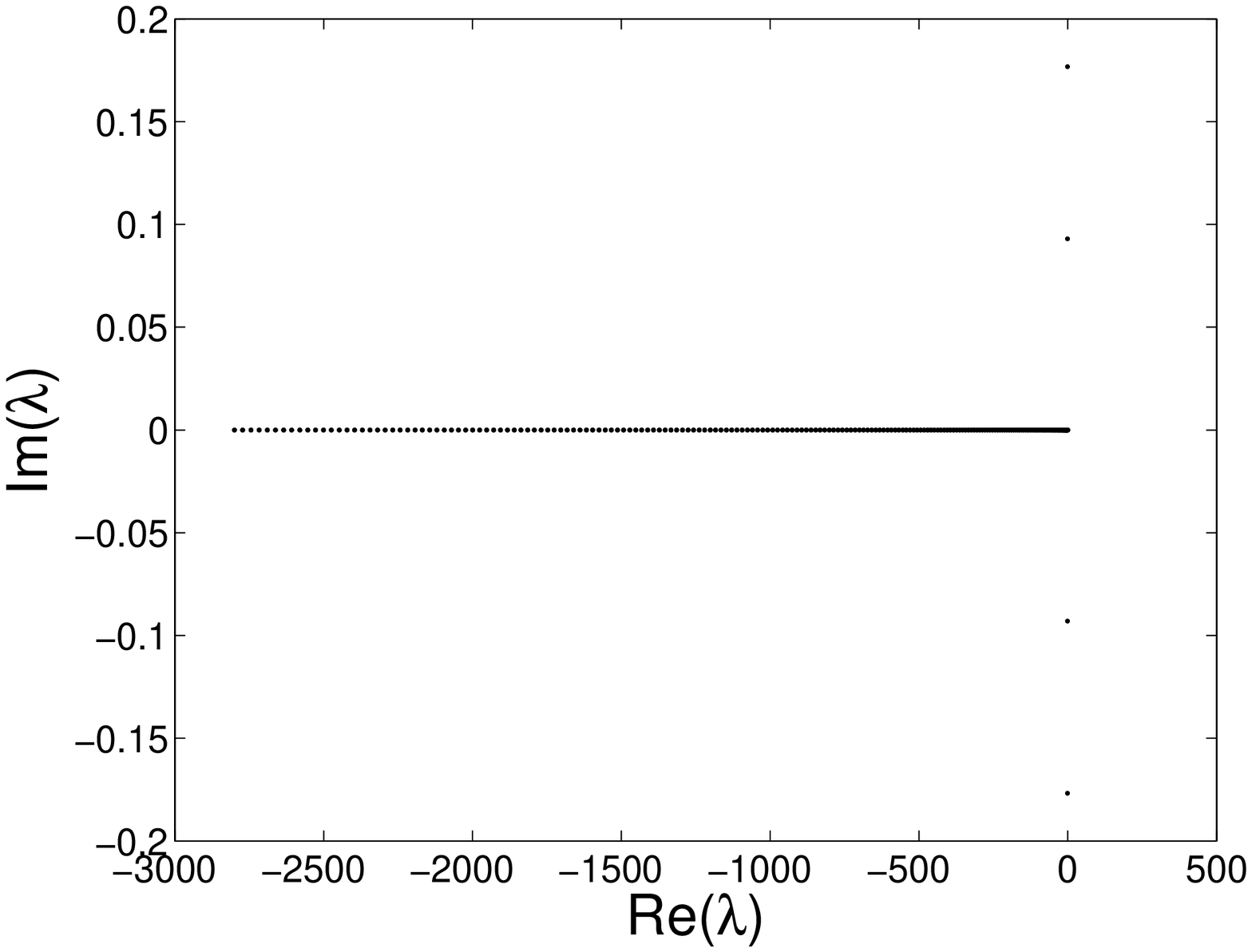}}
\subfigure[$\e =0.03$]{\includegraphics[width=2.3in,height=2.3in]{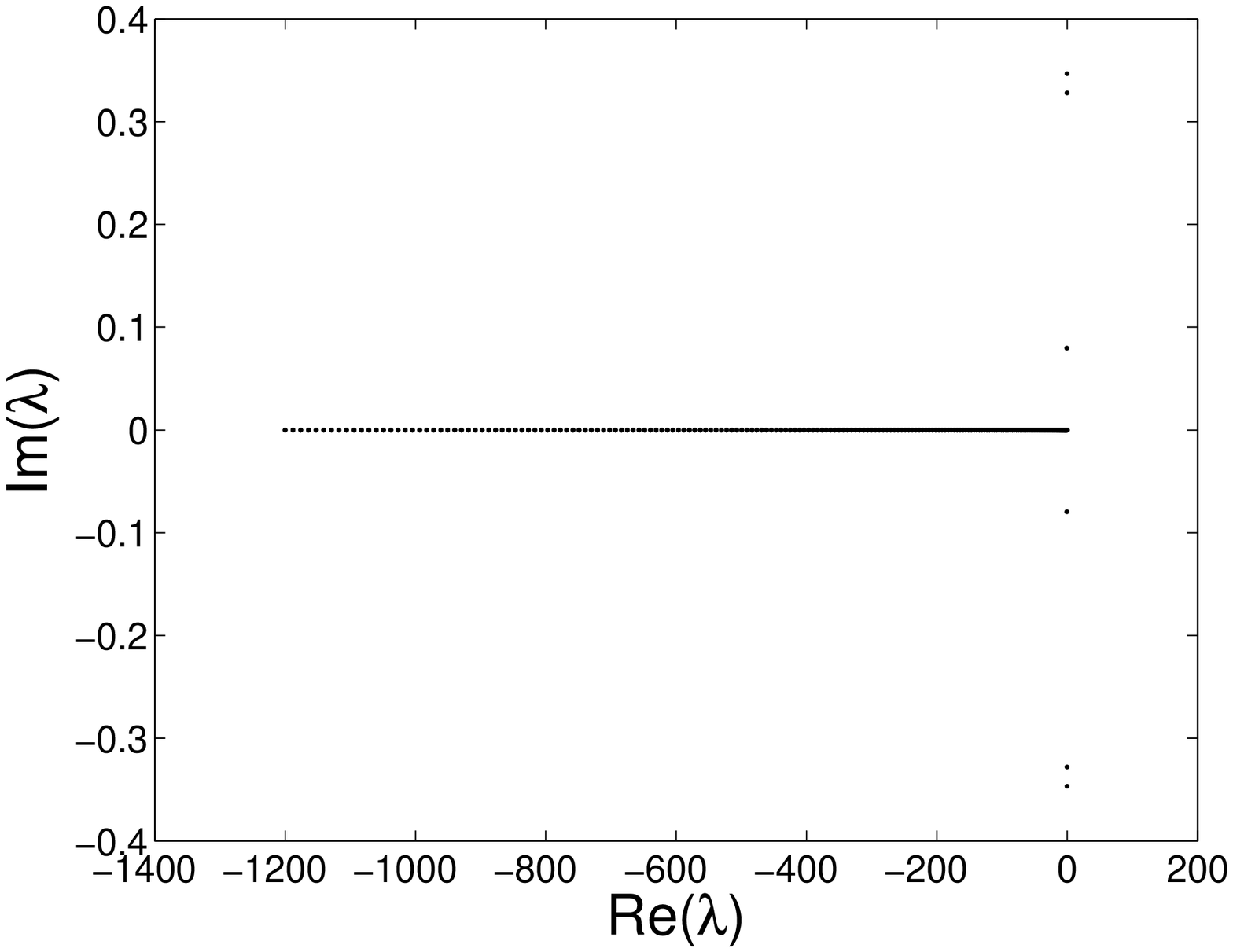}}
\caption{The eigenvalues of the linear system (\ref{le}) when $\hk_1=0$ and $\hk_2=1$, 
$\al =0.7$, and various $\e$.}
\label{ge1-8a}
\end{figure}
\begin{figure}[ht] 
\centering
\subfigure[$\e =0.0004$]{\includegraphics[width=2.3in,height=2.3in]{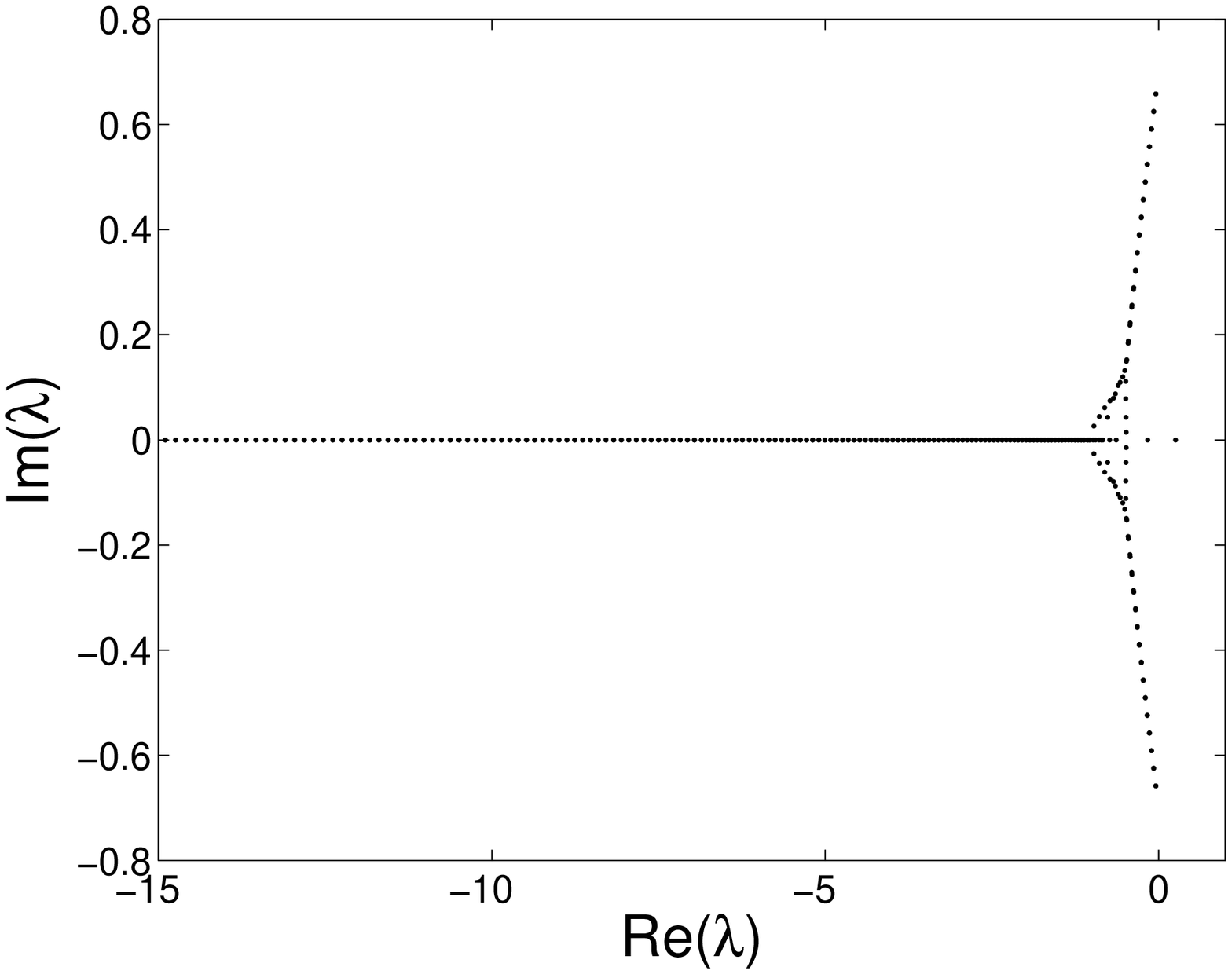}}
\subfigure[$\e =0.00013$]{\includegraphics[width=2.3in,height=2.3in]{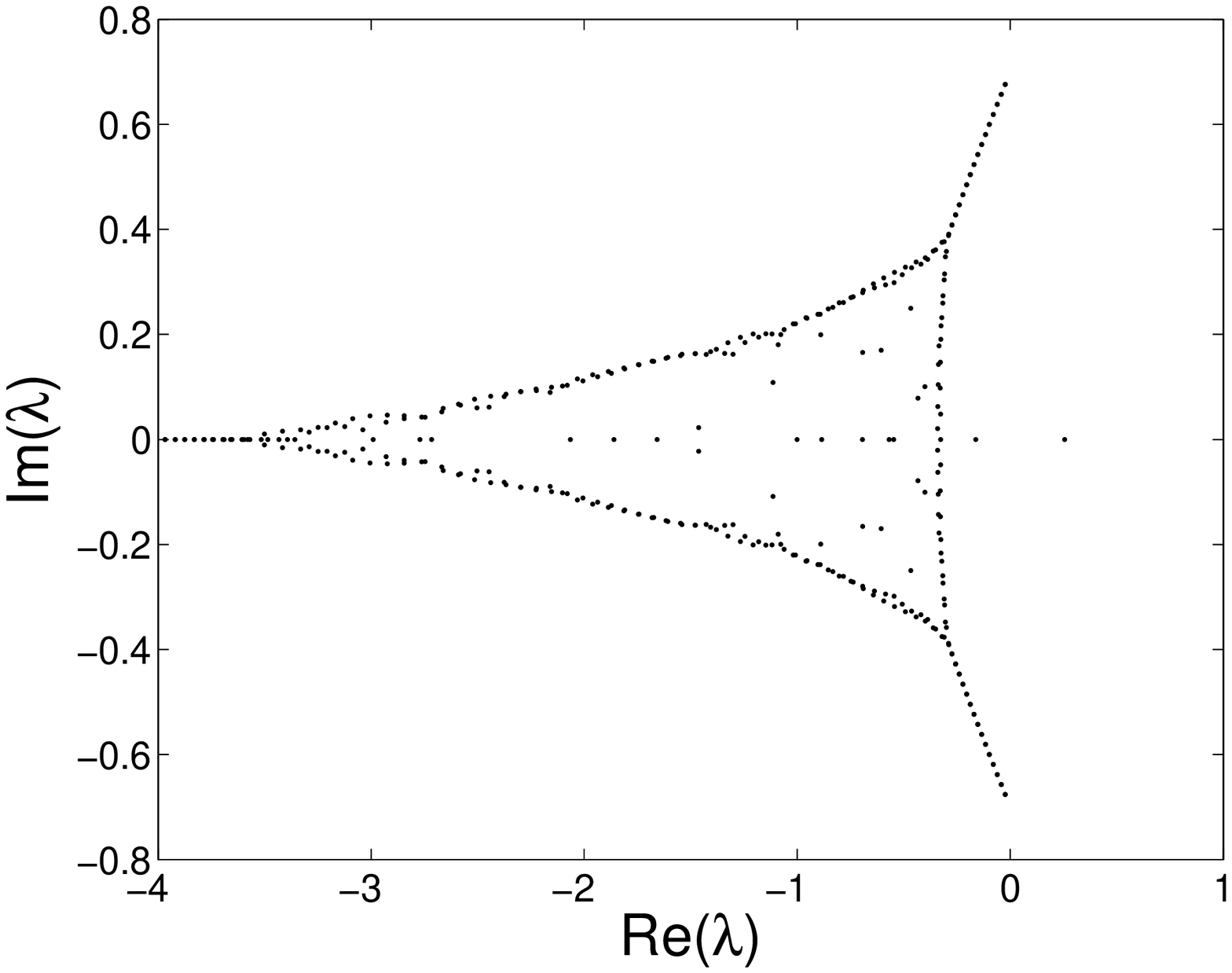}}
\subfigure[$\e \ra 0^+$ limiting picture]{\includegraphics[width=2.3in,height=2.3in]{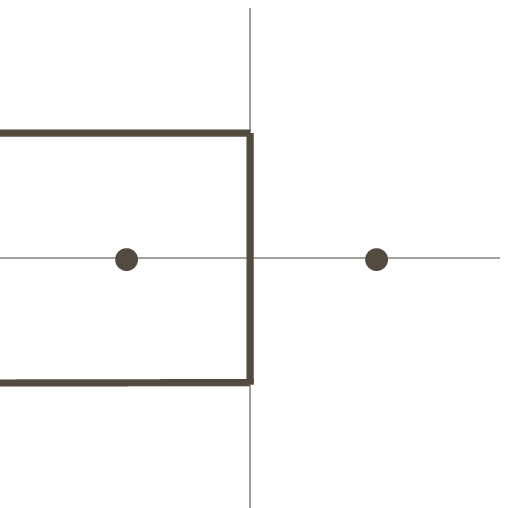}}
\subfigure[$\e =0$]{\includegraphics[width=2.3in,height=2.3in]{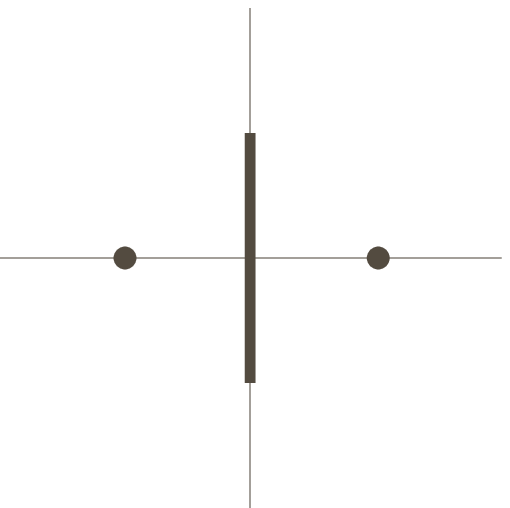}}
\caption{The eigenvalues of the linear system (\ref{le}) when $\hk_1=0$ and $\hk_2=1$, 
$\al =0.7$, and various $\e$ (continued).}
\label{ge1-8b}
\end{figure}
\begin{figure}[ht] 
\centering
\subfigure[$\e =1.5$]{\includegraphics[width=2.3in,height=2.3in]{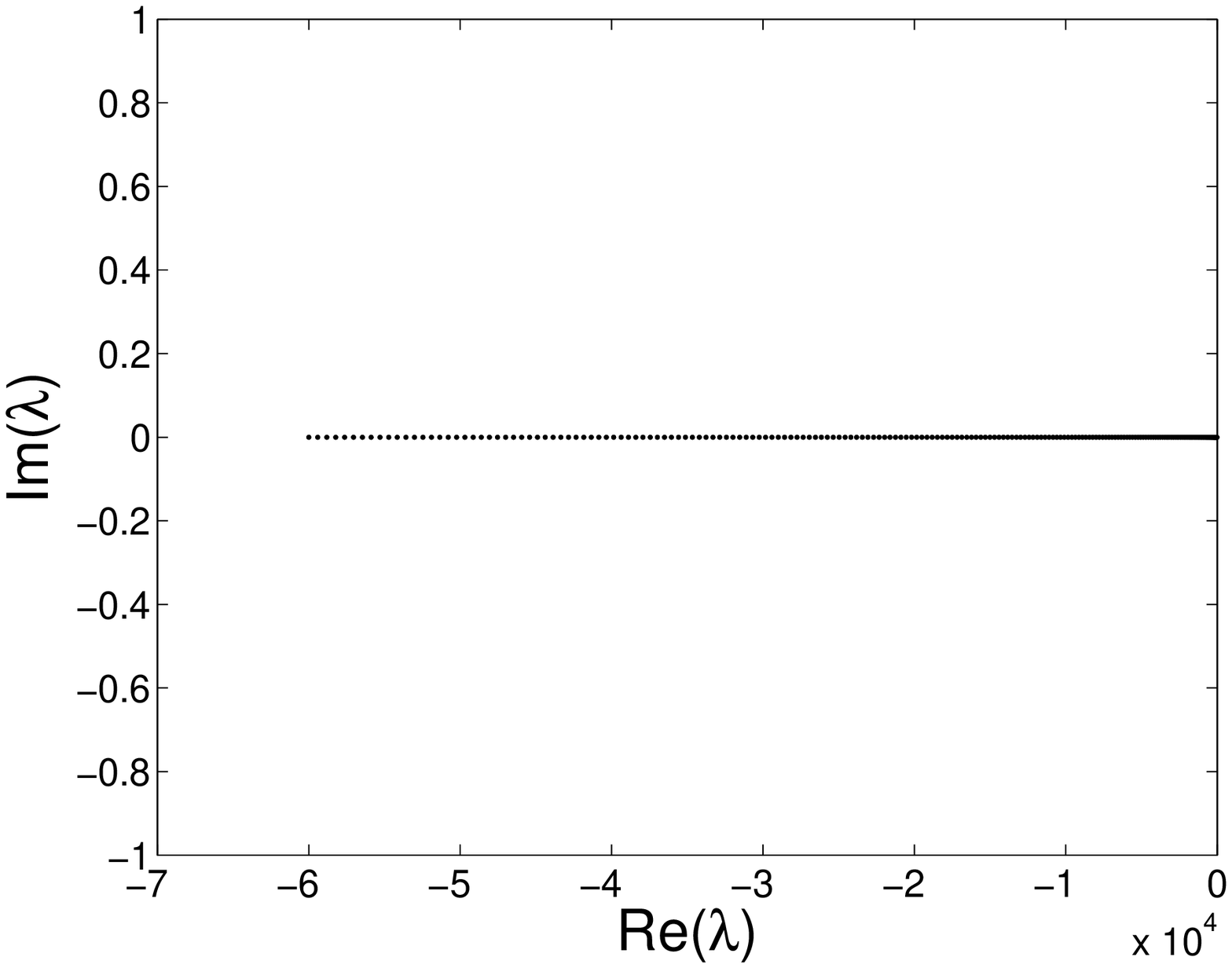}}
\subfigure[$\e =0.00025$]{\includegraphics[width=2.3in,height=2.3in]{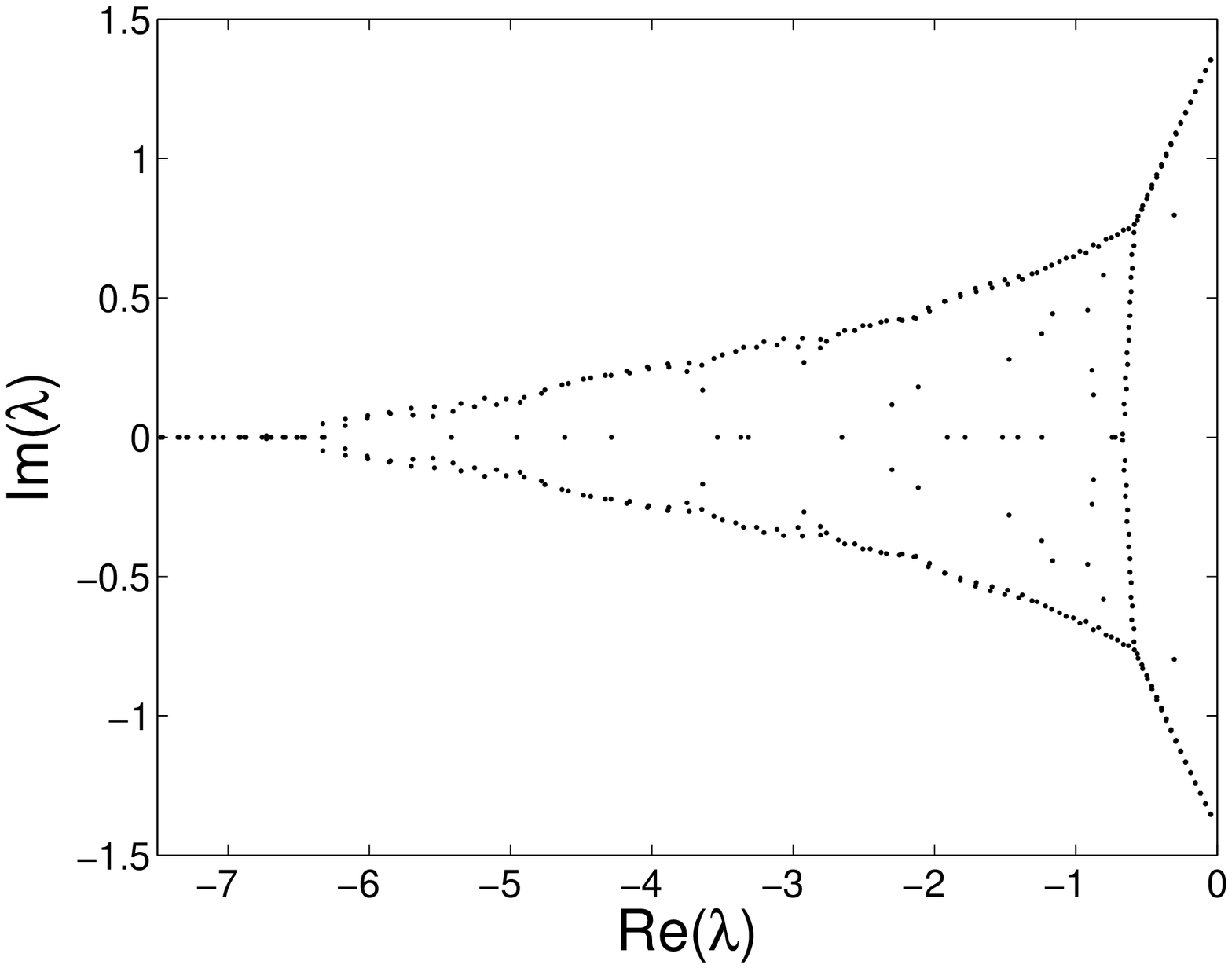}}
\caption{The eigenvalues of the linear system (\ref{le}) when $\hk_1=0$ and $\hk_2=2$, 
$\al =0.7$, and various $\e$.}
\label{ge9-10}
\end{figure}
\begin{figure}[ht] 
\centering
\subfigure[$\e \ra 0^+$ limiting picture]{\includegraphics[width=2.3in,height=2.3in]{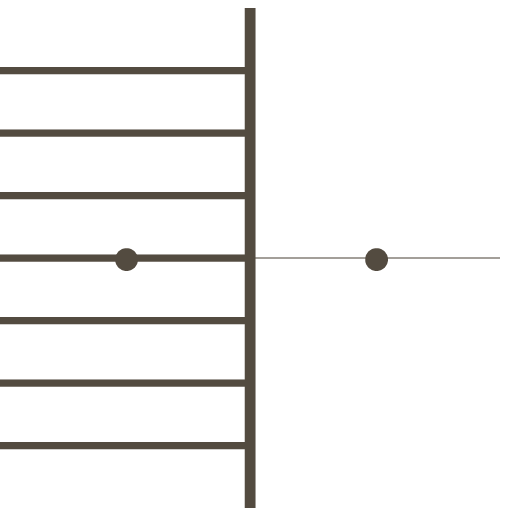}}
\subfigure[$\e =0$]{\includegraphics[width=2.3in,height=2.3in]{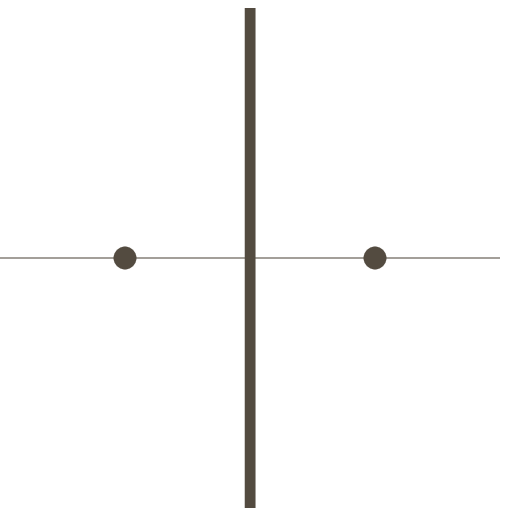}}
\caption{The entire spectrum of the linear NS operator (\ref{le}) when $\al =0.7$, $\e \ra 0^+$
or $\e =0$.}
\label{ge11-12}
\end{figure}

\nid
Figure \ref{ge1-8a}(a) shows the case $\e =0.14$ where there is one positive eigenvalue 
and all the rest eigenvalues are negative. Figure \ref{ge1-8a}(b) shows the case $\e =0.13$ where
a pair of eigenvalues jumps off the real axis and becomes a complex conjugate pair.
Figure \ref{ge1-8a}(c) shows the case $\e =0.07$ where
another pair of eigenvalues jumps off the real axis and becomes a complex conjugate pair.
Figure \ref{ge1-8a}(d) shows the case $\e =0.03$ where
another pair of eigenvalues jumps off the real axis and becomes a complex conjugate pair,
while the former two pairs getting closer to each other. 
Figure \ref{ge1-8b}(a) shows the case $\e =0.0004$ where
many pairs of eigenvalues have jumped off the real axis and a bubble is formed. 
Figure \ref{ge1-8b}(b) shows the case $\e =0.00013$ where the bubble has expanded.
Including many other case testings, our conclusion is that:
As $\e \ra 0^+$, the limiting picture is shown in Figure \ref{ge1-8b}(c). Setting $\e =0$, the 
spectrum of the line $\hk_1=0$ and $\hk_2=1$ of the linear Euler operator has been established 
rigorously (Theorem \ref{PET}) and is shown in 
Figure \ref{ge1-8b}(d), where the segment on the imaginary axis is the continuous spectrum. 
Comparing Figures \ref{ge1-8b}(c) and \ref{ge1-8b}(d), we see that the two eigenvalues represent 
``persistence'', the vertical segment represents ``condensation'', and the two horizontal 
segments represent ``singularity''. Next we study one more line: $\hk_1=0$ and $\hk_2 =2$ 
($\al =0.7$). In this case, there is no unstable eigenvalue. Figure \ref{ge9-10}(a) shows the 
case $\e =1.5$ where all the eigenvalues are negative. As $\e$ is decreased, the eigenvalues 
go through the same process of jumping off the real axis and developing a bubble. 
Figure \ref{ge9-10}(b) shows the case $\e =0.00025$ where the bubble has expanded. As $\e \ra 0^+$, 
the limiting picture is similar to Figure \ref{ge1-8b}(c) except that there is no persistent 
eigenvalue. The cases $\hk_1=0$ and $\hk_2 > 2$ ($\al =0.7$) are all the 
same with the case $\hk_1=0$ and $\hk_2 =2$ ($\al =0.7$). 
Figure \ref{ge11-12}(a) shows the limiting picture of the entire spectrum of the 
linear NS operator as $\e \ra 0^+$. Figure \ref{ge11-12}(b)  shows the entire spectrum of the 
linear Euler operator ($\e=0$) given by Theorem \ref{PET}. 

The fascinating deformation of the spectra as $\e \ra 0^+$ and the limiting spectral picture 
clearly depict the nature of singular limit of the spectra as $\e \ra 0^+$. In the 
``singularity'' part of the limit, there is a discrete set of values for the imaginary 
parts of the eigenvalues, which represent decaying oscillations with a discrete set of 
frequencies. Overall, the ``singularity'' part represents the temporally irreversible 
nature of the $\e \ra 0^+$ limit, in contrast to the reversible nature of the linear 
Euler equation ($\e =0$). Many other fixed points were also studied in \cite{LL08}.

\section{Melnikov Integral}

We propose to use the constant of motion of the 2D Euler equation,
\[
G = \int_{\mathbb{T}^2} \Om^2 dx - \int_{\mathbb{T}^2} |u|^2dx 
\]
to build a Melnikov integral for the corresponding 2D Navier-Stokes equation (\ref{2DNS}).
The goal is to use the Melnikov integral as a measure of chaos, for example, around the line 
of fixed points $\Om = \Ga \cos x_1$ parametrized 
by $\Ga$. $G$ is a linear combination of the kinetic energy and the enstrophy. The 
gradient of $G$ in $\Om$ is given by
\[
\na_\Om G = 2(\Om +\Dl^{-1} \Om )
\]
which is zero along the line of fixed points $\Om = \Ga \cos x_1$ (a basic requirement for the 
convergence of the Melnikov integral). We define the Melnikov integral for the 2D NS (\ref{2DNS}) as 
\begin{eqnarray}
M &=& \frac{\al}{8\pi^2}\int_{-\infty}^{+\infty} \int_{\mathbb{T}^2} \na_\Om G 
[\Dl \Om + f(t,x)] dxdt \non \\
&=& \frac{\al}{4\pi^2}\int_{-\infty}^{+\infty} \int_{\mathbb{T}^2} [\Om +\Dl^{-1} \Om ] 
[\Dl \Om + f(t,x)] dxdt \label{gMel}
\end{eqnarray}
If the heteroclinics conjecture is true, then one can evaluate $\Om$ in the Melnikov integral 
along the heteroclinic cycle. 

As an example, we choose the external force
\begin{equation}
f = a \sin t \cos (x_1 +\al x_2)\ .
\label{EF}
\end{equation}
Then the Melnikov integral (\ref{gMel}) has the expression
\begin{equation}
M = M_1 + a \sqrt{M_2^2+M_3^2} \sin (t_0 +\th )\ ,
\label{12mel}
\end{equation}
where 
\begin{eqnarray*}
& & \sin \th = \frac{M_3}{\sqrt{M_2^2+M_3^2}} \ , \quad \cos \th = \frac{M_2}{\sqrt{M_2^2+M_3^2}} \ , \\
M_1 &=& \frac{\al}{4\pi^2}\int_{-\infty}^{+\infty} \int_0^{2\pi /\al} \int_0^{2\pi}(\Om +\Dl^{-1} \Om )
\Dl \Om \ dx_1 dx_2dt \ , \\
M_2 &=& \frac{\al}{4\pi^2}\int_{-\infty}^{+\infty} \int_0^{2\pi /\al} \int_0^{2\pi}(\Om +\Dl^{-1} \Om )
\cos t \cos (x_1 +\al x_2)  \ dx_1 dx_2dt \ , \\
M_3 &=& \frac{\al}{4\pi^2}\int_{-\infty}^{+\infty} \int_0^{2\pi /\al} \int_0^{2\pi}(\Om +\Dl^{-1} \Om )
\sin t \cos (x_1 +\al x_2)  \ dx_1 dx_2dt \ .
\end{eqnarray*}
For a numerical simulation, we evaluate $\Om (t)$ along the approximate heteroclinic orbit in 
Figure \ref{fi1} with $\Om (0)$ given by 
(\ref{IC}). The time integral is in fact over the interval [$-11.8 ,11.8$] rather than ($-\infty , 
\infty$), which already gives satisfactory accuracy. This is because that $\na G$ decays very 
fast along the approximate heteroclinic orbit in both forward and backward time. Direct numerical 
computation gives that 
\[
M_1 =  -29.0977\ , \quad M_2 = -0.06754695 \ , \quad M_3 = 0 \ .
\]
Setting $M = 0$ in (\ref{12mel}), we obtain that
\[
\sin (t_0 +\pi ) =  \frac{430.77741}{a} \ .
\]
Thus, when 
\begin{equation}
|a| > 430.77741\ ,
\label{MC}
\end{equation}
there are solutions to $M = 0$. The goal is to test whether or not (\ref{MC}) is a criterion for 
the intersection of certain center-unstable and center-stable manifolds. In \cite{LL08}, many 
such tests have been conducted.  

\section{Sommerfeld Paradox}

\begin{figure}[ht]
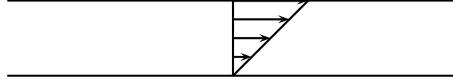

\centering
$$
\psline(-3,0)(3,0)
\psline(-3,1)(3,1)
\psline(0,0)(1,1)
\psline(0,0)(0,1)
\psline{->}(0,0.5)(0.5,0.5)
\psline{->}(0,1)(1,1)
\psline{->}(0,0.25)(0.25,0.25)
\psline{->}(0,0.75)(0.75,0.75)
$$
\caption{Couette flow.}
\label{Cflow}
\end{figure}

The Sommerfeld paradox was originally about Couette flow, in fact, many other fluid flows 
exhibit the same paradox. The Couette flow is between two infinite horizontal planes 
(Figure \ref{Cflow}) where the upper plane moves with unit velocity. The dynamics of 
such a fluid flow is governed by the Navier-Stokes (NS) equations together with a unique 
boundary condition. Specifically 
\begin{equation}
u_{i,t} + u_j u_{i,j} = -p_{,i} +\e u_{i,jj} \ , \quad u_{i,i} = 0 \ ; 
\label{NS}
\end{equation}
defined in the spatial domain $D_\infty = \mathbb{R} \times [0,1] \times \mathbb{R}$, where 
$u_i$ ($i=1,2,3$) are the velocity components, $p$ is the pressure, and $\e$ is the inverse 
of the Reynolds number $\e = 1/R$. The following 
boundary condition identifies the specific flow
\begin{eqnarray}
& & u_1(t, x_1, 0, x_3) = 0 , \quad u_1(t, x_1, 1, x_3) = 1 , \non \\
& & u_i(t, x_1, 0, x_3) = u_i(t, x_1, 1, x_3) = 0 , \ (i=2,3) . \label{nsbc}
\end{eqnarray}

\nid
The classical Couette flow is given by
\begin{equation}
u_1 = x_2, \quad  u_2=u_3=0 . 
\label{Cf}
\end{equation}
In the study \cite{LL08a}, we will focus not only on the neighborhood of the Couette flow (\ref{Cf}), 
but also on the neighborhood of the following sequence of oscillatory shears
\begin{equation}
u_1 = U(x_2) = x_2 + \frac{c}{n} \sin (4n\pi x_2), \ \left ( \frac{1}{2} \frac{1}{4\pi} < c < \frac{1}{4\pi} \right ), \quad  u_2=u_3=0 . 
\label{Os}
\end{equation}

The sequence of oscillatory shears has some remarkable properties:
\[
\lim_{n \ra \infty} \| U(x_2) - x_2 \|_{L^p}  = 0 , \ 
\lim_{n \ra \infty} \| U(x_2) - x_2 \|_{W^{1,p}}  \neq 0 .
\]
That is, in the $L^p$ norms, the oscillatory shears can be regarded as perturbations of the 
Couette shear. One can regard the oscillatory shears as high frequency Fourier monomodes.
In experiments, these oscillatory shears can be regarded as high frequency noises. On the other hand,
in the $W^{1,p}$ norms (i.e. vorticity's $L^p$ norms), these oscillatory shears are not perturbations 
of the Couette shear.

A more precise statement of the Sommerfeld paradox is as follows:
\begin{itemize}
\item Mathematically, the Couette shear is linearly and nonlinear stable for all Reynolds 
number $R$, in fact, all the eigenvalues 
of the Orr-Sommerfeld operator satisfy the bound $\la < -C / R$ \cite{Rom73}.
\item Experimentally, for any $R > 360$ (where $R = \frac{1}{4\e}$ in our 
setting \cite{BTAA95}), there exists a threshold amplitude of
perturbations, of order $\O (R^{-\mu})$ where $1\leq \mu < \frac{21}{4}$ depends on the 
type of the perturbations \cite{KLH94}, 
which leads to turbulence.
\end{itemize}
A mathematically more comfortable re-statement of this experimental claim is as follows: For 
any fixed amplitude 
threshold of perturbations, when $R$ is sufficiently large, turbulence occurs. For any fixed 
$R$, when the 
amplitude of perturbations is sufficiently large, turbulence occurs. There may even be an 
asymptotic relation 
between such amplitude threshold and $R$.

The main idea of our study is as follows \cite{LL08a}: The oscillatory shears (\ref{Os}) 
are perturbations of the Couette shear. 
As $n \ra \infty$, they approach the Couette shear in $L^\infty$ norm. They are linearly 
unstable under the 2D Euler dynamics \cite{LL08a}. This leads to the existence of an unstable 
eigenvalue 
of the corresponding viscous Orr-Sommerfeld operator, when the Reynolds number $R$ is 
sufficiently large \cite{LL08a}. Notice that these 
oscillatory shears are not fixed points anymore under the Navier-Stokes dynamics. Nevertheless, 
they only drift very slowly. 
The important fact is that here the unstable eigenvalue of the Orr-Sommerfeld operator is 
order $\O (1)$ with respect to 
$\e = 1/R$ as $\e \ra 0^+$. Intuitively, this fact should lead to relatively long finite time nonlinear 
growth near the oscillatory shears (and the Couette shear). Such a growth will manifest 
itself in experiments as transient 
turbulence. Here the amplitude 
of the perturbation from the Couette shear will be measured by the deviation of the 
oscillatory shears from the Couette shear and 
the perturbation on top of the oscillatory shears.

\section{Future Directions and Open Problems}

In terms of analytical studies on chaos in fluids, the promising future direction will be 
the periodic boundary condition model where analysis can benefit from Fourier analysis. 
The heteroclinics conjecture is a challenging open problem on top of the open problem 
on the existence of invariant manifolds for 2D Euler equation. On the other hand, Galerkin 
truncations and numerical simulations all indicate that such a heteroclinics should exist. 
The current numerical capability is still limited in dealing such simulations. Of course, 
future improvement of numerical capability is always a hope. The question of how useful 
Melnikov integrals can be for predicting chaos in fluids, is still unclear. Once again, the 
first obstacle is the open problem on the existence of invariant manifolds for 2D Euler equation.
2D Euler equation may not have any invariant manifold at all. In principle, Melnikov integrals
serve as good predicting tools for chaos only when the phase space structures are very special in the 
sense that the Melnikov prediction on the intersection of certain center-unstable and 
center-stable manifolds leads to the existence of certain homoclinics or heteroclinics. 
Here for the 2D Navier-Stokes equation, the phase space structures are not clear at all, 
for example, what are the fixed points or limit cycles? how close they are from the simple 
fixed point we are studying?

The phase space structures of Couette flow, Poiseuille flow, and plane Poiseuille flow etc. 
also drew a lot of attentions recently. Most of the works are numerical. Due to the limitation 
of the numerical capability, analytical results are very precious. Analytically, the norms of 
the phase space play a fundamental role which is often missing from numerical simulations. 
It may be possible to have a better picture of a small region in the phase space by feedback 
studies between analysis and numerics. It may also be possible to prove the existence of 
certain nontrivial fixed points or limit cycles. In finite dimensional dissipative systems,
chaos is often associated with certain homoclinics or heteroclinics. It is not clear whether or 
not this is still true for fluid flows with boundary layers. There are some positive indications 
from recent numerical simulations \cite{HGCV08}.

\end{document}